\documentclass[12pt,letterpaper]{article}
\usepackage{amsmath,amssymb,pgf,pgfarrows,pgfnodes,float,appendix, hyperref}
\usepackage{graphicx}
\usepackage{subfigure}
\usepackage[margin=0.9in]{geometry}
\usepackage{slashed}
\newcommand{\be}{\begin{equation}}
\newcommand{\ee}{\end{equation}}
\newcommand{\bea}{\begin{eqnarray}}
\newcommand{\eea}{\end{eqnarray}}

\title{{\rm\footnotesize \qquad \qquad \qquad \qquad \qquad \ \qquad \qquad \qquad \ \ \ \ \ \                      RUNHETC-2014-06  SCIPP 14/05}\vskip.5in     The Super BMS Algebra, Scattering and Holography}
\author{Tom Banks\\
Department of Physics and SCIPP\\
University of California, Santa Cruz, CA 95064\\
{\it and}\\
Department of Physics and NHETC\\
Rutgers University, Piscataway, NJ 08854\\
E-mail: \href{mailto:banks@scipp.ucsc.edu}{banks@scipp.ucsc.edu}
\\
}
\date{}
\begin{document}
\maketitle

\begin{abstract}
I propose that the proper framework for gravitational scattering theory is the representation theory of the super-BMS algebra of Awada, Gibbons and Shaw\cite{ags}, and its generalizations.  Certain representation spaces of these algebras generalize the Fock space of massless particles.  The algebra is realized in terms of operator valued measures on the momentum space dual to null infinity, and particles correspond to smearing these measures with delta functions.  I conjecture that scattering amplitudes defined in terms of characteristic measures on finite spherical caps, the analog of Sterman-Weinberg {\it jets}\cite{stermanweinberg}, will have no infrared (IR) divergences.  An important role is played by singular functions concentrated at zero momentum, and I argue that the formalism of Holographic Space-Time is the appropriate regulator for the singularities.  It involves a choice of a time-like trajectory in Minkowski space.  The condition that physics be independent of this choice of trajectory is a strong constraint on the scattering matrix.  Poincare invariance of $S$ is a particular consequence of this constraint.  I briefly sketch the modifications of the formalism, which are necessary for dealing with massive particles. I also sketch how it should generalize to AdS space-time, and in particular show that the fuzzy spinor cutoff of HST implements the UV/IR correspondence of AdS/CFT.
\end{abstract}

\section{Introduction}

The attempt to construct a quantum theory of gravitation began with seminal work of DeWitt, Feynman, Mandelstam and Weinberg in the 1960s \cite{dwfmw} .  These authors showed that the scattering theory of massless spin two particles in Minkowski space led uniquely to the Einstein Lagrangian as a low energy effective description of the scattering amplitudes, when we impose the constraints of Poincare invariance, unitarity, and analyticity/crossing symmetry.  These results are enough to guarantee that string theory, a non-field theoretic formulation of quantum gravity, is well approximated at low energy by an effective field theory.

There are two problems with this approach to the quantum theory of gravity.  The first is that we have found it impossible to construct S matrix theories of quantum gravity, which do not have exact super-Poincare invariance\footnote{In perturbative string theory, this is related to instabilities of scalar fields with either tachyonic or flat potentials.  However, even those String Landscape enthusiasts who believe that there is no correlation between the size of the c.c. and the breaking of SUSY, balk at claiming that there will be Poincare invariant scattering theories in Minkowski space, which are not super-Poincare invariant.} .  The second is that gravitational scattering amplitudes, at least in low enough dimensions, are beset with IR divergences.  While soft-graviton theorems are enough to ensure the existence of finite inclusive cross sections\cite{weinv1}, there has been very little work\cite{kulakhoury} generalizing the Fadeev-Kulish definition of IR finite {\it amplitudes} in QED to theories of gravitation.

In this paper, I will argue that the first problem solves the second.  That is, the assumption of super-Poincare invariance of gravitational scattering, suggests the introduction of a new framework for that scattering theory based on the super-BMS algebra of Awada Gibbons and Shaw\cite{ags}, in which IR finite amplitudes for the analog of Sterman-Weinberg {\it jets} are the natural objects of study.  I will formulate the super-BMS algebra in a language dual to that of \cite{ags}, working on the momentum space dual to null infinity, and also suggest that their algebra has to be enriched to take into account quantum numbers other than spin and momentum.  The dual momentum space is simply a null cone $P^2 = 0$ and the positive and negative energy branches of the cone correspond to outgoing and incoming particles respectively.  There are separate copies of the algebra $\psi_{\alpha}^{\pm} (P)$ defined on the individual branches, and the scattering matrix intertwines them
$$S \psi_{\alpha}^+ (P) = \psi_{\alpha}^- (P) S .$$  The generators of the algebra are operator valued ``half-measures"\footnote{This is my name for linear functionals on the space of half densities on the null cone. Half densities are spinors, square roots of differential forms.}

It turns out that the $P = 0$ mode of $\psi_{\alpha} (P)$ is crucial to much of what follows.  This is a somewhat singular concept, since it retains angular information about a sphere that has shrunk to a point.  This is less surprising if we think in terms of null infinity itself.  Null infinity is not a manifold, because its two components are glued together by an infinite sphere at space-like infinity.  The zero momentum point is dual to this infinite sphere.  It is the focus of both IR divergences, and the low energy theorems that help to control them. Strominger\cite{andy} has recently emphasized that continuity conditions at space-like infinity are crucial to formulating gravitational scattering theory, and that the S matrix formula relating past and future BMS generators is, when written in a (possibly singular) particle basis, the origin of Weinberg's soft graviton theorems.  He views these generators as spontaneously broken symmetries.  I prefer to think of the super-BMS generators as an algebra that generates an IR finite space of asymptotic states.  In this picture there is no unique vacuum state, but rather a generic state of zero momentum generators serves as the vacuum, while scattering states have some of the zero momentum generators vanishing, in a kind of ``jet isolation".  I'm sure there is some way in which the two points of view can be reconciled.

The zero momentum/infinite spatial sphere singularities are regularized by thinking of null infinity as the limit of the null boundary of a nested sequence of causal diamonds with finite area holographic screens.  Specification of a complete sequence of such nested screens, with areas going from the Planck area to infinity, is equivalent to the choice of a time-like trajectory through space-time.
The combination of the Holographic Principle and causality tells us that the Hamiltonian describing evolution along that trajectory must be time dependent.  At each finite time, it must split the Hilbert space into tensor factors, ${\cal H}_{in} \otimes {\cal H}_{out}$, such that ${\cal H}_{in}$ is finite dimensional, with dimension $$D_{{\cal H}_{in}} = e^{A/4} ,$$ with $A$ the area in Planck units of the screen of the diamond corresponding to that time interval.  ${\cal H}_{out}$ is infinite dimensional, but we can regularize this by simply thinking of a very large proper time interval $N L_P$, restricting to the diamond for that time interval and then taking the limit $N \rightarrow\infty$.  In this paper, we will restrict attention to geodesic trajectories. 

For finite $N$, the physics obviously depends on the choice of trajectory.  For example, for a given choice of trajectory, and $N$, there will be scattering events which occur outside the causal diamond of that trajectory over the proper time interval $[- N , N] L_P$.  We must certainly insist that this dependence go away as $N$ goes to infinity.  Applying this principle to the set of geodesics in Minkowski space, we obtain Poincare invariance of the S matrix as a corollary of the requirement of cutoff/trajectory independence. We will discuss a more formal statement of this principle below.

For finite $N$, a natural choice of variables is to retain the super-BMS generators up to some cutoff, and a natural choice of cutoff is defined in terms of the spectrum of the Dirac operator on the holographic screen.  Indeed, we will see that the super-BMS generators are sections of the spinor bundle over the sphere at infinity.  This prescription also leads us to a natural tensor factorization when we consider nested causal diamonds.  The smaller diamond will have a smaller area, corresponding to a lower cutoff on the Dirac operator on the sphere.   Since a Dirac cutoff is the same as an angular momentum cutoff, any rotation invariant choice of commutation relations for the super-BMS variables will define a natural tensor factorization of the Hilbert space of the diamond of size $N$ into variables associated with a diamond of smaller size, $n$, and the remainder.

The rest of this paper is organized as follows:  in the next section we define the super-BMS algebra and discuss generalizations of it.  We also define scattering states and low energy theorems for emission of zero energy quanta.  This discussion will be brief, and rely mostly on Strominger's work. In section 3 we introduce the regularized version of the algebra and the formalism of HST.  We discuss the implications for compact dimensions of space.  We also provide a formal statement of the principle of trajectory independence, which has implications at finite $N$, and can provide a definition of a quantum theory of space-times whose asymptotic boundaries are quite different from those of Minkowski space.  In section 4, we give a brief discussion of the way to incorporate massive particles into our scattering theory.  In Section 5, we briefly mention the modifications necessary to deal with AdS space, and the connection to the UV/IR connection in AdS/CFT. Section 6 is devoted to Conclusions and directions for further research.

\section{Scattering Representations of the super-BMS Algebra}

The conformal boundary of $d$-dimensional Minkowski space consists of two cones each of which has a conformal structure with one null direction, $u_{\pm}$.  The fibers transverse to the null direction are conformal to round $d - 2$ spheres.  The conformal group, $SO(1, d - 1)$ of the sphere acts on the null coordinate via $u_{\pm} \rightarrow e^{\zeta (\omega_{\mu\nu} ; \Omega)} u_{\pm}$, in such a way that the vector fields
$$P_{\mu} = (1, {\bf \Omega}) \partial_u , $$ defined in the conformal frame with a round metric, transform as a $d$-vector under $SO(1, d - 1)$.  Here ${\bf \Omega}^2 = 1$ parametrizes a point on the sphere, and $\omega_{\mu\nu}$ are the parameters of the Lorentz transformations. For fixed conformal frame, the area of the spheres increases from $0$ to $\infty$ as $u$ does.  The two cones are joined along their infinite area spheres, which is space-like infinity.

At each point on null infinity, with finite $u_{\pm}$, there are unique outward pointing and inward pointing positive energy null normals.  We define the sign of energy such that outward going energy is positive, so that incoming momenta have negative energies.  Outward and inward pointing refer to the orientation of the spatial momentum normal to the $d - 2$ sphere.  The momentum space dual to null infinity is thus a light cone in momentum space: the locus of points with $P^2 = 0$.  The positive branch of this cone corresponds to future null infinity and the negative branch refers to past null infinity.  The point $P = 0$, which is common to the two cones is dual to space-like infinity.  For the future cone, the points are labelled by $P (1, {\bf \Omega})$, where ${\bf \Omega}$ is a unit three vector of positive orientation, and $P \geq 0$.  Similarly, a point on the past cone is $- P (1, \bf \Omega)$.  Two points $P,Q$ on (say) the future cone, point in the same direction if and only if $P\cdot Q = 0$.  They then have the form $$P_{\mu} = P (1 , {\bf \Omega}) ,$$
$$Q_{\mu} = Q (1 , {\bf \Omega}) ,$$ and we define $$M_{\mu} (P, Q) \equiv {\rm min}\ [P,Q] (1, {\bf \Omega}) .$$

The minimal super-BMS algebra in any number of dimensions is
$$ [ Q_{\alpha} (P) , \bar{Q}_{\beta} (Q) ]_+  = \gamma^{\mu}_{\alpha\beta} M_{\mu} (P,Q) \delta (P \cdot Q ) .$$  Here $Q_{\alpha}$ is a minimal spinor, in the sense that all possible Majorana and Weyl conditions have been imposed.  $\bar{Q}_{\beta} = Q^{\dagger\alpha} \gamma^0_{\alpha\beta}$, is its Dirac conjugate.  The commutation relations are compatible with the constraint
$$ \gamma^{\mu}_{\alpha\beta} P_{\mu} Q_{\beta} (P) = 0 .$$ We impose this constraint and note that it makes $Q_{\alpha}$ into a section of the spinor bundle over the sphere.  A perhaps more familiar  way of saying this is that the constraint says that the generator is a null plane spinor in the local tangent plane to the sphere.  At fixed $P$, the representation space of this algebra is the space of states of a massless super-particle.  We know that in space-time dimension greater than $11$, there cannot be a consistent non-trivial scattering matrix for such particles.

More general super-BMS algebras exist.  The simplest are those based on extended super-symmetries, which can often be thought of as coming from compactifications of an eleven dimensional theory on a compact manifold that preserves some SUSY.  In such a scenario, null infinity is the product of null infinity in some $d < 11$ and a compact manifold ${\cal K}$.  The spinor bundle is a tensor product of the spinor bundle over the $d - 2$ sphere and that over ${\cal K}$.  Expanding in a complete set of eigenspinors of the Dirac operator\footnote{This will be replaced by the Dirac operator coupled to background fluxes, for flux compactifications.} on ${\cal K}$, labelled by an integer $n$, we have $$Q_A (P_{11}) \rightarrow Q_{\alpha\ a} (P, n) ,$$ and we can write a super-algebra
$$[Q_{\alpha\ a} (P, n) , \bar{Q}_{\beta\ b} (Q, m) ]_+ = \gamma^{\mu}_{\alpha\beta} M_{\mu} (P,Q) \delta (P \cdot Q ) \sum_k Z_{(s^1 \ldots s^k)} (n,m) \Gamma^{(s^1 \ldots s^k)}_{ab}.$$ The sum runs over all $k$ forms on ${\cal K}$.  

If the manifold has covariantly constant spinors, these are zero modes of the Dirac algebra, and give corresponding minimal copies of the SUSY algebra at fixed $P$\footnote{The same goes through {\it mutatis mutandis} for flux compactifications.}.  Our hypothesis is that there must be at least a minimal SUSY algebra.  We should also insist that the full algebra be large enough to {\it ensure} the existence of graviton states in the representation at fixed $P$.  Note that this does not require there to be an extended SUSY algebra or larger than minimal SUSY.  We have been careful to refrain from saying that our algebra has anything to do with symmetries (spontaneously broken or otherwise) of the S matrix, and indeed we have not yet defined the S matrix.  

To fully specify the algebra, we must specify commutation relations for the $Z$ operators with the spinor generators and with themselves, satisfying the graded Jacobi identities. For maximally SUSic compactifications, we expect the $Z's$ to be central charges, corresponding to wrapped branes.  With half-maximal SUSY or less the $Z$'s can be non-abelian, and some of the fermionic generators will be charged under the non-abelian algebra.  I conjecture that the kinematic classification of finite dimensional super-algebras, which have a representation containing exactly one graviton, is the analog of finding super-Poincare invariant compactifications of string/M-theory.  This conjecture is far from proof, but if true its combination with the holographic principle implies that the continuous moduli of semi-classical string/M theory are actually discrete. Once we look at a finite causal diamond, we will find that the internal algebra must have a finite dimensional unitary representation.

The collinear singularity in the anti-commutation relations implies that the operators $Q_{\alpha} (P)$ are distributions of some sort.  We will in fact assume that they are operator valued measures, which must be smeared with measurable functions to give operators
$$ Q [f] = \int dP Q_{\alpha} (P) f^{\alpha} (P) .$$ We define Sterman-Weinberg jet\cite{stermanweinberg} states by the equation
$$ Q[f] | Jet \rangle = 0, $$ unless $f$ has support at non-zero $P$, only in a finite set of spherical caps, centered around vectors $P_k$.
We must be careful about the singular point of the cone, $P=0$.  As $P$ goes to zero, we insist that $f (P)$ vanish in a set of annuli surrounding each of the caps, but it can remain a finite function of ${\bf \Omega}$ outside those annuli.   

A more precise mathematical characterization of these generators will probably identify them with {\it half measures}\footnote{This is related to the familiar mathematical characterization of wave functions as half densities, but my name is funnier. The functions $f_{\alpha}$ are half densities.}, since their anti-commutator is a measure.   The important point is that particle states correspond to smearing the super-BMS generators with delta functions, and are {\it a priori} singular.  

There are a couple of important points to be noted before we deal with a crucially important subtlety of the formalism.  First, Jets, or their singular particle limits, are characterized as excitations of a field operator, so that fact that permutation of particles/jets is a gauge symmetry is an automatic consequence of the formalism. Secondly, the fact that the spinor generators satisfy {\it anti}-commutation relations enforces the usual connection between spin and statistics. We will define the S matrix as a unitary intertwining map between the representation spaces of the past and future super-BMS algebras:

$$S Q_{\alpha}^- [P,m] = Q_{\alpha}^+ [P,m] S .$$ The action of $S$ on the space of scattering states, will take one set of jets, described by a function $f_{\alpha}^- (P, n)$, into another, characterized by $f_{\alpha}^+ (P,n)$\footnote{One interesting aspect of this is that momentum conservation is not guaranteed by any of these equations. As Strominger shows, the S matrix relation between the past and future BMS generators {\it does not} imply that these are symmetries of the S-matrix, but only soft graviton theorems.  In conventional field theoretic discussion of graviton scattering, momentum conservation is built in, but a more axiomatic discussion, based on the super-BMS algebra, would have to impose it as an extra constraint. In recent investigations of such S-matrix constructions\cite{amplitudhedron}, momentum conservation, expressed in terms of spinor variables, is a constructive principle, which leads to analytic forms for the amplitudes.}. 
Strominger\cite{andy} has shown that this equation is valid for the bosonic BMS sub-algebra of the minimal super-BMS algebra in pure classical gravitational scattering theory on Christodoulu-Klainerman spaces. He and his collaborators have shown how it is related to Weinberg's soft graviton theorems. A crucial role in that discussion is played by the zero momentum states, which we have neglected to discuss.

Our definition of a scattering representation of the algebra, must deal with the singular limits $P \rightarrow 0$ of the super-BMS generators, both in the past and the future.  For these generators, the definitions of the signs of energy, as well as the orientation of the spatial component of the momentum is ambiguous.  Strominger's discussion shows that we must include {\it both} signs, both in the past and the future, in order for the above S matrix equation to be consistent with non-trivial scattering.  Otherwise, the incoming energy at each solid angle on the sphere would have to be exactly equal to the outgoing energy at the anti-podal angle.  The zero momentum generators have to re-arrange themselves (which means in particular that they cannot be represented by zero on the scattering states) to allow for scattering.  

We define the scattering representation of the algebra by insisting only that the zero momentum generators vanish in small annuli surrounding the initial and final jets.  An annulus on a $d - 2$ sphere is the disjunction of two concentric spherical caps.   The full past and future scattering representations are direct sums of Hilbert spaces with different numbers of incoming (outgoing) jets.   The vanishing of the zero momentum generators on annuli defined by the jets is called {\it jet isolation}.  

The fact that we have to represent the infinite dimensional algebra of zero momentum generators, parametrized by measurable functions on the sphere vanishing in each jet sector on annuli of as yet unspecified width, makes the mathematical formulation of this theory ambiguous.  In the next section we will regularize that ambiguity by introducing finite causal diamonds.

\section{Regularization of the super-BMS Algebra With Finite Causal Diamonds}

The Holographic Principle provides a way to regularize the zero momentum and co-linear singularities of the super-BMS algebra.  According to this principle\cite{tHfsb}, the Hilbert space associated with a causal diamond with a finite area\footnote{We use the word area, in $d$ dimensional space-time, to denote the $d-2$ volume of a space-like $d - 2$ surface.} holographic screen is the exponential of one quarter of the area, in Planck units.  It's natural to associate this finite dimensional Hilbert space with a representation of a finite dimensional deformation of the super-BMS algebra.  In the earliest papers on Holographic Space-time, Fischler and I introduced variables that were motivated by the fuzzy two sphere. They were elements of the spinor bundles over the fuzzy two sphere, $N \times N + 1$ matrices.  Conventional fuzzy geometry is restricted to manifolds with Kahler structures, and cannot preserve the rotation group of spheres of dimension higher than two.

More recently, Kehayias and I\cite{tbjk} introduced a cutoff on the Dirac operator on the spinor bundle of the holographic screen, as a covariant definition of fuzzy geometry, which was capable of preserving symmetry, as well as some of the topological information of the limiting continuous space.  If we apply that cutoff to a $d - 2$ dimensional sphere, we find a number of spinors parametrized by an anti-symmetric $d - 2$ form
$$\psi_a^{(m_1 \ldots m_{d-2})} (n) .$$ each index runs from $1$ to $N$, where, following the holographic relation between area and entropy, $N$ is the radius of the sphere in Planck units.  The last statement depends on the assumption that the representation space of the quantum operators corresponding to the Dirac eigenfunctions, is generated by acting with these operators on a space of $N$ independent dimension (probably a single state).  The label $n$ represents an eigen-basis of the Dirac operator on the internal manifold. It has a finite range, representing the fact that the internal manifold has a finite size in Planck units.   The operator algebra is
$$[\psi_a^{(m_1 \ldots m_{d-2})} (n), \psi_b^{\dagger\ (k_1 \ldots k_{d-2})} (p)]_+ = \delta{ab}\delta^{(m_1 \ldots m_{d-2}) , (k_1 \ldots k_{d-2})} Z_{np} , $$ plus commutation relations for the $Z_{np}$, which make this into a finite dimensional super-algebra, with representation generated by the action of the $\psi^{\dagger}$'s on a state of vanishing $\psi$'s .  The detailed structure of this super-algebra is the Holographic Space Time analog of the information contained in the geometry of compactifications in string theory.  The details of this correspondence have not yet been fully worked out, even for the simplest compactifications.   However, it's clear from the general formalism that if these ideas are right then continuous moduli are artifacts of length scales much larger than the Planck scale.

The absolutely simplest compactification is one for which $n$ and $p$ are just the indices on a spinor in $11 - d$ dimensions and $Z_{np}$ is a c -number Kronecker delta.  This is maximally supersymmetric compactification on a ``minimal size" torus.
For this compactification, a Hamiltonian describing proper time evolution along geodesics has been proposed. More precisely, in references\cite{scatt} a whole class of Hamiltonians was proposed, parametrized by a polynomial of order $N^{d-4}$.   

One first constructs the matrix 
$$M_m^k =  \psi_a^{\dagger (k_1 \ldots k_{d-3} k)}\psi_{a\ (m k_1 \ldots k_{d-3})}  . $$   If we think of these as numerical matrices and tensors, and picture $\psi$ and $\psi^{\dagger}$ as $d - 2$ cubes, then $M$ is diagonal when the spinors vanish except on blocks on the main diagonal of the hypercube.  We define jet states to be those which are annihilated by all of those off diagonal $\psi's$ and for which the state is a tensor product in the Hilbert spaces of the individual non-vanishing blocks.  We have some number of blocks of size $1 \ll K_i \ll N$, and one block of size $N - \sum K_i$.  The interaction term in the Hamiltonian takes the form
$$ V = \frac{1}{N^{2(d - 3)}} {\rm Tr}\ P(M) ,$$ where $P$ is a polynomial of order $\sim N^{d-4}$ whose coefficients have finite 
large $N$ limits.  This Hamiltonian is invariant under a large unitary group, acting independently on the indices of $\psi$.  It has an $SO(d - 1)$ subgroup, under which these variables transform like spinor spherical harmonics.  The block diagonal decomposition breaks the large symmetry to the product of block symmetries.  We associate this breaking with the localization of states on the sphere.   That is, each block of size $K_i$ represents functions concentrated near a different point ${\bf \Omega_i}$ on the sphere.  The localization is of order $\frac{1}{K_i}$ in each angular direction.  The vanishing variables represent the annuli from our discussion in the previous section.   The full Hamiltonian in a fixed jet sector is
$$ \sum P_{0\ i} + V,$$ where the $P_{0 i}$ are the energy operators appearing in the emerging super-BMS algebra as $K_i \rightarrow \infty$.  One can show that the ratios of energies are proportional to the ratios of $K_i^{d-3}$.  To describe time evolution along a trajectory one identifies a subset of variables associated with a nested sequence of causal diamonds, and prescribes a time dependent Hamiltonian $H_{in} (n)$ identical to the one above, with $N \rightarrow n$ for each $n < N$.
The rest of the variables at time $n$ are described by a Hamiltonian $H_{out} (n)$, which is supposed to be determined in order to satisfy the following consistency condition:  

\begin{itemize}

\item We introduce an infinite set of other quantum systems, corresponding to other geodesics in Minkowski space. This is done by prescribing the overlaps between causal diamonds along one trajectory and those along another.  That is, we identify a tensor factorization of each Hilbert space into the overlap\footnote{By overlap we mean the maximal area causal diamond in the overlap of the two diamonds as sets.} and something else.  On each such overlap, the initial states and time evolutions along each trajectory define a density matrix.  The two density matrices must agree, up to a unitary transformation.

\item Along a given trajectory some event might happen in a given causal diamond.   If an event along some other trajectory is space-like separated from the event in question, then, using DOF that only appear in a larger causal diamond along the first trajectory, we can introduce commuting degrees of freedom, and a term in $H_{out}$ 
of the first trajectory, to copy the dynamics along the second, thus satisfying the consistency conditions in a simple fashion.

\item When events are close together, this simple procedure is insufficient and we must be more careful about solving the consistency conditions.

\item The consistency conditions guarantee Poincare invariance of the S matrix.

\end{itemize}

For all of these models (without worrying about the short distance consistency conditions) one can show the following\cite{scatt}

\begin{itemize}

\item There is a scattering theory.  That is, an incoming jet state will evolve into a superposition of outgoing jet states.  The total energy
$E = \sum K_i^{d-3}$ is conserved, although the individual $K_i$ and the number of jets, are not.

\item Up to a point, one can track the jets into smaller and smaller causal diamonds, but the number of jets coming out of some small causal diamond can be different than the number of incoming jets.
The consistency conditions for sufficiently large separations between causal diamonds allow us to tie together these vertices into time ordered Feynman like diagrams.   The interactions are concentrated in small causal diamonds because of the explicit factors of $n^{ - 2(d-3)}$ in front of the interaction Hamiltonian.  This is the origin of space-time locality in these models.

\item At large impact parameter, the scattering is dominated not by contributions from Feynman diagrams, but by interactions mediated by the large block of variables, of size $N - \sum K_i$.  These variables contribute zero energy at large $N$.   They are to be identified with the zero momentum super-BMS generators of the previous section.  The amplitude scales with energy and impact parameter like that in an eikonal approximation to graviton exchange\footnote{Here we use language appropriate to a covariant gauge in perturbative gravity. In a physical gauge, this amplitude comes from explicit non-local terms in the Lagrangian.}. Incidentally, this shows that the power of $N$ in front of the polynomial interaction could not fall off any more slowly than $N^{-2{d-3}}$, since the S matrix would not exist.  Furthermore, if we place a higher inverse power of $N$ in front of this term then one can show that other powers of the energy would be conserved, and the S-matrix would be unity.

\item  When $E = \sum K_i^{d -3} \sim n^{d-3}$, in a causal diamond of size $n \gg 1$, the Feynman diagram picture breaks down because the number of constrained and unconstrained variables is similar.  If the Hamiltonian is a fast scrambler\cite{sekinosusskind}, the entire system will begin to equilibrate in a time of order $n {\rm ln}\ n$.
Once it is equilibrated, the probability of finding a particle state of energy $E$ is of order $e^{- n E}$, which indicates that the state is thermal with temperature of order $\frac{1}{n}$ .  The equilibrium state, by the Stefan Boltzmann law, is meta-stable, with a lifetime of order $n^{d - 1} \gg n$.  The process cannot be captured by a diagram with local vertices.

\end{itemize}

The conclusion is that all of these models have qualitative properties similar to those associated with a quantum theory of gravity.   However, satisfying the consistency conditions exactly puts constraints on the coefficients in the polynomial.  Among these is Lorentz invariance of the S matrix.

\section{Massive Particles}

For massive particles, the momentum space representation of the conformal boundary of Minkowski space must be amended.  Time-like infinity is a singular point in the traditional conformal compactification 
of ${\cal M}^d$.  To describe massive particles, we have to assign a full $d-2$ sphere to this point.  Furthermore, again as a consequence of the singularity, there is no unique orientation for the null vector at some point on this sphere.  Instead, the mass parametrizes the correct linear combination of $P (1, \pm {\bf \Omega})$ describing the outgoing momentum.  Thus
$$P_{\mu} (m) = P(1, {\bf \Omega}) + \frac{m^2}{2P} (1, - {\bf \Omega}) \equiv P_{\mu}^+ ({\bf \Omega}) + m^2 P_{\mu}^- ({\bf \Omega}), $$
with $2 P^2 > m^2$ is a unique representation of an outgoing positive energy massive momentum in terms of a pair of positive energy null vectors with opposite orientation.  

This decomposition allows us to build operators representing massive particle jets from the super-BMS algebra we have already constructed.  More precisely, in our construction of the positive energy generators, we made a choice and constructed operators
$ Q_{\alpha}^{++} (P_{\mu})$, which anti-commute to null vectors with outward pointing spatial component ${\bf \Omega}$, at $P_{\mu}^+ = P (1,{\bf \Omega})$, rather than inward pointing ${\bf - \Omega}$.  There is another set of generators, $Q_{\alpha}^{+-}$,obtained by a parity flip, which anti-commute to $P_{\mu}^-$ instead\footnote{Note that the normalization is chosen so that $P_{\mu}^+ P^{\mu\ -} = 1$.   This makes $P_{\mu}^-$ singular when $P$ goes to zero, but we only use it for massive particles, where $2 P^2 > m^2$, so this causes no problems.}.   It is also clear that $[Q^{++} , Q^{+-} ]_+ = 0$.

Massive BPS particles come from representations of a SUSY algebra
$$ [Q_{\alpha} ,  \bar{Q}_{\beta} ]_+ = \gamma^{\mu}_{\alpha\beta} P_{\mu} - M \delta_{\alpha\beta} ,$$ where we've suppressed all other labels on the generators, and $M$ is an operator on some space of BPS states.  The momentum is on-shell, which means that $\slashed{p} = m$ for one of the eigenvalues of $M$.  The $Q$ operators, for each mass satisfy the equation
$$ (\slashed{p} + m)Q = 0. $$  Using the decomposition of a massive momentum in terms of oppositely oriented null momenta,  We can solve this equation as
$$ Q = Q^{++} (P_{\mu}^+ ) + M Q^{+-} (P_{\mu}^+),  $$ which anti-commutes to the massive momentum.  Thus, we can describe BPS states in terms of the super-BMS algebra, if we know their mass.

Knowledgable readers will immediate think about massive stable non-BPS states\cite{senetal}.  I believe that a discussion of such states will involve considerations of unitarity of the S-matrix.  A glimpse of what is involved will be apparent after we deal with BPS black holes of arbitrary charge.   If we accept the contention that, for fixed momentum, the super-BMS algebra has a finite dimensional representation, then the BPS mass matrix $M$ must be a finite dimensional, and therefore bounded operator.  At first glance this seems to contradict well known results from string theory, where the BPS charges are obviously unbounded, because they are generators of $U(1)$ gauge groups.  However, for any value of moduli, fixed in Planck units, only a finite number of these charged BPS states have masses below the Planck mass.  I conjecture that the part of the spectrum described by the super-BMS representation, only includes these sub-Planckian states.   On the other hand, using models of the type discussed in the previous section, it is clear that we can make charged states of arbitrary charge via collisions\footnote{This is of course obvious from general physical principles, but I wanted to emphasize that the explicit quantum mechanical models of the previous section will obey this principle, once we identify bounded charges describing "elementary" BPS states.}.  They will be stable composites of the hypothesized elementary BPS states described in the representation space of the  super-BMS algebra.  We will discover them by imposing unitarity on the S matrix, and finding that it is not unitary in the original Hilbert space, without adding bound states.

A similar tactic should work for the stable states which are not BPS.
As far as I know, all known states of this type can be thought of as bound states of BPS particles of opposite charge, which carry a non-trivial charge under a finite gauge group.  In perturbative string theory language, this is the K-theory charge\cite{senetal} of the vector bundles representing collections of oppositely charged D-branes.
Again, we could discover these states from a failure of unitarity in the original Hilbert space.  

In a way, this is similar to what happens for orbifolds in perturbative string theory.  One makes a projection on the perturbative spectrum and gets an S-matrix satisfying tree-unitarity, but this fails at one loop level and we discover twisted states, which carry a representation of a new finite gauge group.

It's clear that this discussion is very preliminary.  A lot more work is needed to understand how stable massive states fit into a definition of scattering theory in terms of representations of the super-BMS algebra.

\section{Super BMS $\rightarrow$ Super AdS}

It is interesting to consider limits of the variables describing finite causal diamonds, which correspond to $AdS_d$ space, rather than Minkowski space.   The causal structure of the two is quite different, because massless particles can travel out to the boundary and return to a time-like trajectory in the interior, in finite proper time, of order the AdS radius.  The conformal boundary of the space-time is thus a cylinder, $R \times S^{d-2}$, and our parametrization of the super-BMS algebra in terms of angles and longitudinal momenta fails.  The boundary causal diamonds still have infinite area holographic screens, so we expect the variables of HST to approach operator valued distributions on the boundary.  The conformal algebra of the boundary is $so (2, d-1)$ and there are a finite number of super-conformal extensions of it, which could be limits of the constant modes of the super-algebra of HST variables.   The black hole spectrum in $AdS_d$ space behaves like the spectrum of a $d - 1$ dimensional CFT, so, following \cite{maldagkpw}, we are motivated to identify the quantum theory on the boundary with a quantum field theory.

It's then clear that the localized generators of the pixel algebra of HST should converge to an algebra containing the super-conformal current algebra on the boundary.  Positivity of the metric in Hilbert space forces this algebra to contain Schwinger terms, so that there are no states of the limiting theory annihilated by the local generators.  

The UV/IR connection of AdS/CFT actually follows from the prescription of \cite{tbjk} for describing fuzzy compact manifolds.  A finite volume in Planck units corresponds to a finite UV (angular momentum) cutoff on the sphere, and we are proposing that our operators converge to quantum field operators on the sphere.  Note however that a local causal diamond cannot correspond simply to an angular momentum cutoff on the boundary CFT.  The boundary CFT contains unbounded operators even with an angular momentum cutoff.  These must be realized as limits of a finite number of fermion operators.  Field theories are universal fixed points of the renormalization group, and can be realized as limits of many different kinds of cutoff system\footnote{For example, continuum $\phi^4$ field theory in 2 and 3 dimensions is the limit both of the Ising model, and of cutoff $\phi^4$ field theory.  Only one of these has bounded operators and a finite dimensional Hilbert space in finite volume. The cutoff of field theory by finite causal diamonds in HST, is analogous to the Ising cutoff.}.  Clearly there's much left to be understood in connecting HST to AdS/CFT.   

\section{Conclusions}  

The super-BMS algebra provides a plausible arena for an infrared finite gravitational S matrix in Minkowski space of dimensions between $4$ and $11$.  Its detailed mathematical formulation is obscured by the necessity of including an infinite number of zero momentum generators.  It should be emphasized that these degrees of freedom are, as shown by Strominger, related to soft gravitons in the conventional Fock space formulation of the scattering problem.
The real trick is to understand the extent to which they decouple from conventional amplitudes.

In this paper we've provided a regularization of the super-BMS algebra, via the formalism of HST.  The zero momentum DOF are clearly identified in finite causal diamonds, and are responsible for black hole entropy, and large impact parameter scattering, at least in the explicit Hamiltonian matrix models introduced in \cite{scatt} .  They play an important role in finite diamonds, but decouple from the S matrix for Sterman-Weinberg jets, which should be finite.  

There are a number of important directions for future work on the super-BMS algebra:  \begin{itemize} \item Probably the most important is to begin a classification of non-trivial extensions of the algebra, and their relation to compactifications of string theory.  This is the way to investigate the conjecture that moduli are actually discrete.

\item One should understand the relation between the point of view taken here, and Strominger's view of BMS generators as spontaneously broken symmetries.  One intriguing question raised by Strominger's point of view is whether one can, after all, have quantum gravitational models with broken SUSY in asymptotically flat space. The way to investigate this is to study the super-BMS algebra for a field theory model of SUGRA coupled to a Goldstino.

\item Much more work is needed to flesh out our ideas about the representation of massive particles, and the relation of the super-BMS algebra to superconformal current algebra in AdS limits of HST.
\end{itemize}

Of course, it will also be necessary to find parameters that satisfy the consistency conditions of HST and give a Lorentz invariant S-matrix.
I hope to return to these questions in further publications, but hope even more strongly that this paper will motivate others to investigate these questions.

\vskip.3in
\begin{center}
{\bf Acknowledgments }
\end{center}
 I would like to thank W.Fischler for numerous conversations about the material in this article, and A. Strominger for conversations about the BMS algebra.  This work was supported in part by the Department of Energy.

\end{document}